\documentclass[draft]{agujournal2019}
\pdfoutput=1
\usepackage{color}
\usepackage{graphicx}
\usepackage{amsmath}
\usepackage{amssymb}
\usepackage{cancel}
\graphicspath{{./figs/}}
\makeatletter
\def\input@path{{./figs/}}
\makeatother

\renewcommand{\vec}[1]{\mathbf{#1}}

\newcommand{\comment}[1]{}
\newcommand{\ig}[2]{\includegraphics[width = #1]{#2}}

\journalname{Geophysical Research Letters}

\begin{document}
\title{Electron-scale reconnection in three-dimensional shock turbulence} 
\authors{J. Ng\affil{1,2}, L.-J. Chen\affil{2}, N. Bessho \affil{1,2}, J.Shuster \affil{1,2}, B. Burkholder \affil{2,3}, J. Yoo\affil{4} }
\affiliation{1}{Department of Astronomy, University of Maryland, College Park, MD 20742, USA}
\affiliation{2}{NASA Goddard Space Flight Center, Greenbelt, MD 20771, USA}
\affiliation{3}{University of Maryland, Baltimore County, MD 21250, USA}
\affiliation{4}{Princeton Plasma Physics Laboratory, Princeton, NJ 08540, USA}
\correspondingauthor{Jonathan Ng}{jonng@umd.edu}
\date{\today}

\begin{keypoints}
\item Turbulent reconnection events in a 3D quasi-parallel shock simulation  are characterized.
\item Both strong and weak guide field events are found, consistent with the range of values seen in observations. 
\item Reconnection sites have different 3D orientations not captured by 2D simulations. 
\end{keypoints}

\begin{abstract}
  Magnetic reconnection has been observed in the transition region of quasi-parallel shocks. In this work, the particle-in-cell method is used to simulate three-dimensional reconnection in a quasi-parallel shock. The shock transition region is turbulent, leading to the formation of reconnecting current sheets with various orientations. Two reconnection sites with weak and strong guide fields are studied, and it is shown that reconnection is fast and transient. Reconnection sites are characterized using diagnostics including electron flows and magnetic flux transport. In contrast to two-dimensional simulations, weak guide field reconnection is realized. Furthermore, the current sheets in these events form in a direction almost perpendicular to those found in two-dimensional simulations, where the reconnection geometry is constrained.
\end{abstract}

\section{Introduction}

Magnetic reconnection is a process in which magnetic field lines in a plasma change their topology, often accompanied by the conversion of stored magnetic energy to kinetic energy of accelerated particles \cite{priest:2000,yamada:2011}. Reconnection plays an important role in laboratory and space  plasma processes, including sawtooth crashes in tokamaks \cite{goeler:1974}, magnetic substorms in the Earth's magnetosphere \cite{angelopoulos:2008} and  solar flares \cite{sweet:1969}. 

A recent development in the study of reconnection has been the observation of purely electron-scale reconnection regions, in which ions do not participate in the reconnection process \cite{phan:2018, gingell:2019}. This differs from the standard picture of collisionless reconnection \cite{ishizawa:2004,malyshkin:2008,yamada:2011}, where the reconnection region consists of an electron-scale layer within a wider ion-scale diffusion region. These electron-only reconnection regions have been observed in the Earth's magnetosheath and shock transition region \cite{phan:2018,gingell:2019}, foreshock \cite{wang:2020foreshock,liu:2020}, magnetotail \cite{lu:2020} and laboratory experiments \cite{shi:2022}.

In this work we focus on reconnection in the turbulent environment downstream and in the transition region of quasi-parallel shocks. Numerous electron-scale current sheets are observed in these regions \cite{gingell:2019,retino:2007}, leading to a favourable environment for electron-scale reconnection \cite{gingell:2017,gingell:2019,pyakurel:2019}. While existing kinetic simulation studies in the shock environment have shown reconnection at both electron and ion scales, the simulations have been two-dimensional \cite{bessho:2018shock, bessho:2020, bessho:2022,lu:2021}, restricting reconnection to a single plane. These studies find  strong guide-field reconnection, in which there is a magnetic field component parallel to the current. In contrast, observations show a wide range of guide fields \cite{gingell:2019}, with statistics showing that  weak guide-field events are slightly less favoured, but still observed. Three-dimensional effects have also been shown to enhance the reconnection rate in laminar electron-scale current sheets \cite{pyakurel:2021}. It is therefore important to study shock-driven reconnection in three dimensions.

We perform fully-kinetic, three-dimensional simulations of a high Mach number quasi-parallel shock. We find numerous reconnecting and non-reconnecting current sheets, and demonstrate reconnection in two events with strong and weak guide fields.

\section{Simulation}

We perform three-dimensional simulations of a quasi-parallel shock using the fully-kinetic particle-in-cell code VPIC \cite{bowers:2008,bowers:2008proc}. The initial condition consists of a uniform plasma and electromagnetic fields, with $B_x = B_0 \cos \theta$, $B_y = B_0 \sin \theta$ and $E_z = -V_{flow} B_0 \sin\theta$. The initial plasma moves in the negative $x$ direction with velocity $-V_{flow}$. The lower $x$ boundary uses conducting walls for fields and reflecting walls for particles, while plasma and the $z$-component of the electric field are injected at the upper $x$ boundary with the initial field and flow values. The $y$ and $z$ boundaries are periodic. The simulation domain is $2000\times 500\times 200$ $(d_e)^3$ covered by $4000 \times 1000 \times 400$ cells, and is initialized with 150 particles per species per cell (note that typical 2D simulations take place in what we define as the $x$-$y$ plane). Physical parameters used in the simulation are $\omega_{pe}/\omega_{ce} = 4$, $m_i/m_e = 100$, $\beta_e = \beta_i = \sqrt{2}$, $\theta = 30^\circ$ and $M_A = 10$. Here $\omega_{pe}$ is the electron plasma frequency, $\omega_{ce}$ the electron cyclotron frequency, $\beta$ the ratio between thermal pressure and magnetic pressure for either species and $M_A = V_{flow}/v_A$ the Alfv\'en Mach number of the injected plasma. As the simulation develops, the shock front propagates from the lower $x$ boundary in the positive $x$ direction. Unless otherwise mentioned in the text, length scales in the paper are normalized to $d_e$, and velocities to $c$, and number densities to the initial upstream density. 

\begin{figure}
    \centering
    \ig{3.375in}{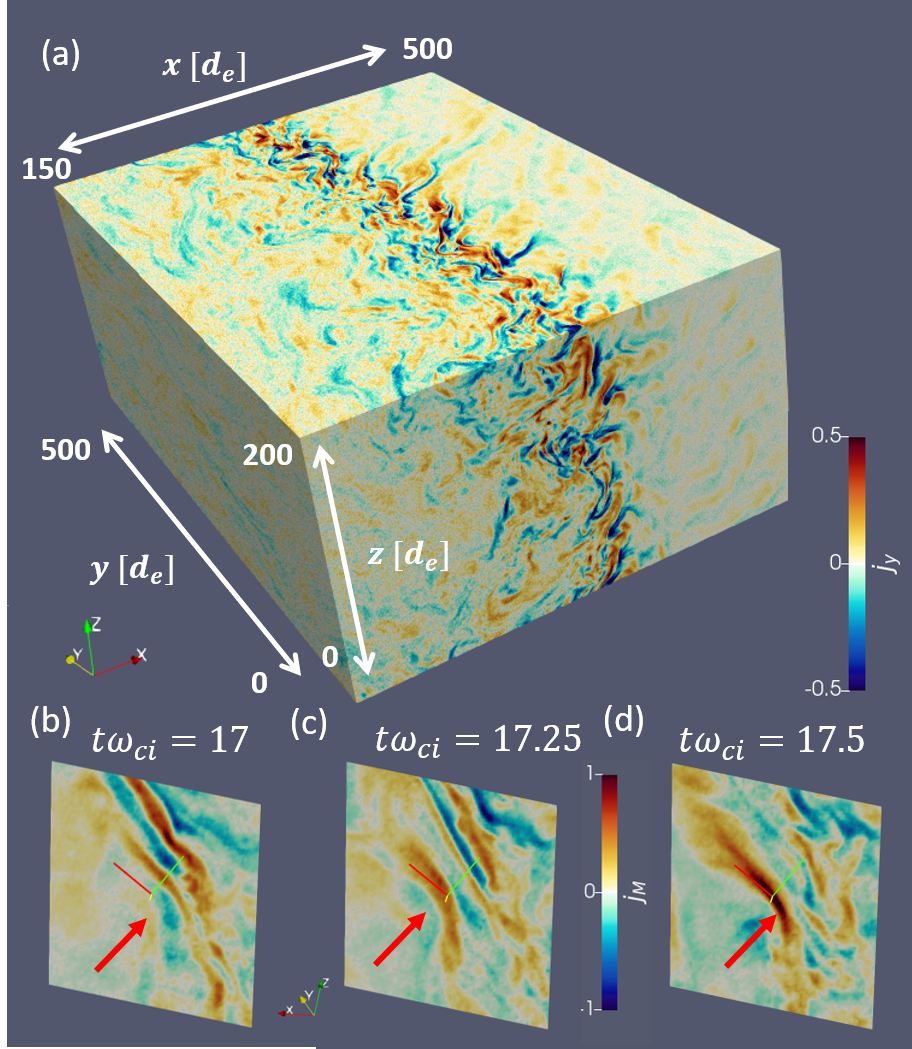}
    \caption{(a) Three-dimensional overview of the shock transition region at $t\omega_{ci} = 17.5$. The upstream region is on the right ($+x$), and plasma flows  in the $-x$ direction, while the shock propagates in the $+x$ direction. Multiple regions with intense $j_y$ are visible in the $x$-$y$ and $x$-$z$ planes. (b-d) Evolution of a current sheet. Colored panels show the current density in the out-of-plane direction, while the arrows point out the structure that evolves into the reconnecting current sheet. The red, yellow (out-of-plane) and green lines on the figure show the $L$, $M$ and $N$ directions where $L$ is the direction of the reconnecting field, $M$ is in the direction of the current and $N$ is normal to the current sheet. These lines have length 10 $d_e$. The simulation axes are shown by a glyph at the bottom.}
    \label{fig:overall}
\end{figure}

\section{Results}

The shock propagates in the positive $x$ direction at 1.9 $v_A$, giving rise to an Alfv\'en Mach number of 11.9 in the laboratory frame. An overview of the shock transition region at $t\omega_{ci} = 17.5$ is shown in  Fig.~\ref{fig:overall}. In the quasi-parallel shock geometry, the interaction between incident and reflected particles in the foreshock leads to strong electromagnetic fluctuations and plasma turbulence, resulting in the formation of numerous current sheets in the transition region and downstream of the shock. Examples of current sheets can be found in Fig.~\ref{fig:overall}(a), in which regions of enhanced positive and negative $j_y$ can be seen. As can be seen in Fig.~\ref{fig:overall}(a), where there is structure in both the $x$-$y$ and $x$-$z$ planes, the three-dimensional geometry allows current sheets, and hence reconnection, to have different orientations.

\begin{figure}
    \centering
    \ig{3.375in}{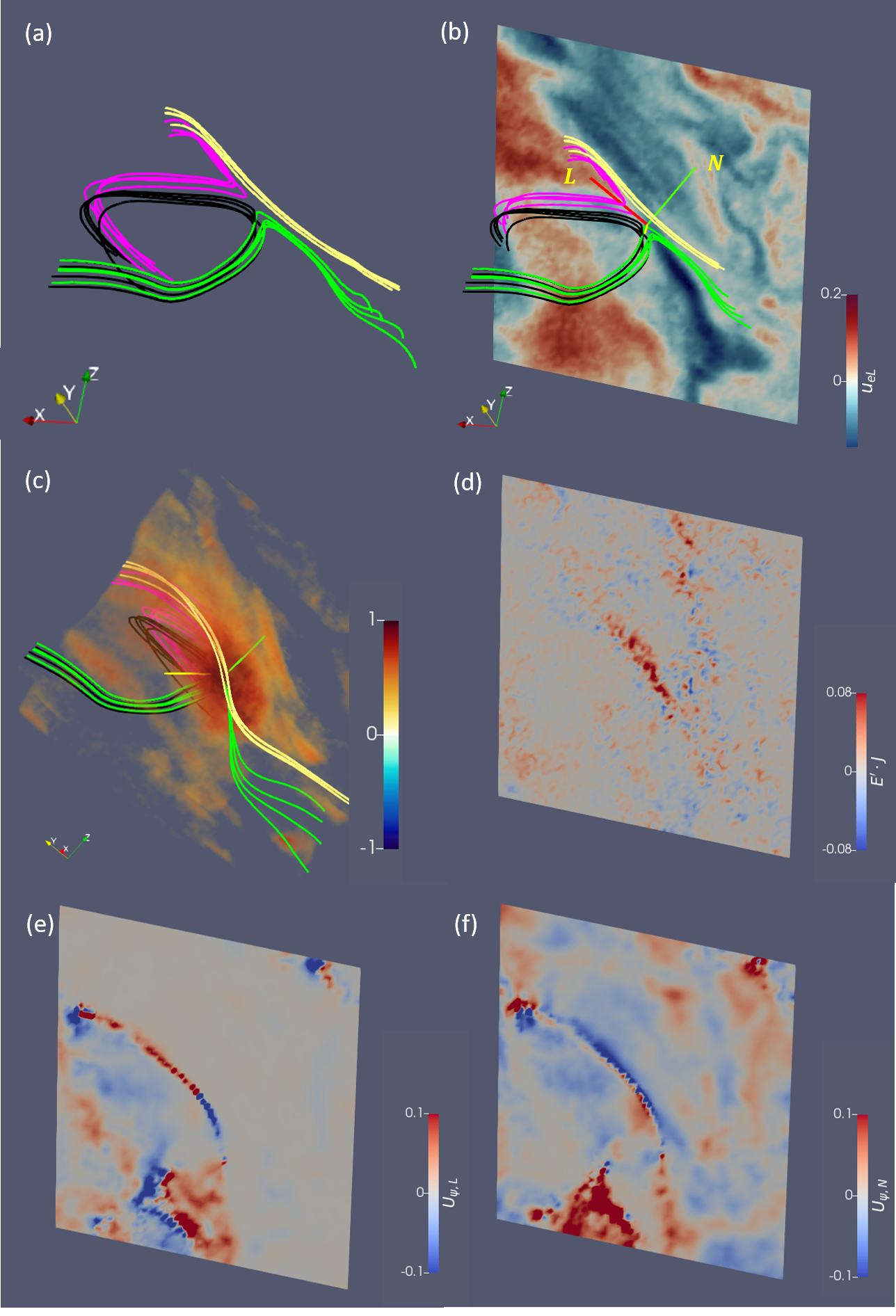}
    \caption{Structure of the reconnection region from various angles. (a) Magnetic field line configuration with colors showing the different topologies associated with reconnection. Black and yellow are associated with the inflow, while pink and green are associated with the outflow regions. (b) Electron velocity $u_{eL}$ in the $L$-$N$ plane. The red, yellow and green lines in the figure show the $L$, $M$ and $N$ directions respectively. $L$ and $N$ are labeled and $M$ is approximately out of plane. The $L$, $M$ and $N$ lines have length 10 $d_e$. (c) Three-dimensional volume rendering of current density $j_M$ showing the reconnection region and field lines. (d) Energy conversion from fields to particles $\smash{\vec{E}'\cdot\vec{J}}$. (e,f) Magnetic flux transport (see text) diagnostic showing active reconnection. }
    \label{fig:site}
\end{figure}

Similar to prior two-dimensional simulations of quasi-parallel shocks \cite{bessho:2018shock,bessho:2020}, both reconnecting and non-reconnecting current sheets are found in the simulation. Here we focus on two active reconnection sites, the first of which is shown in Figs.~\ref{fig:overall}(b-d) and \ref{fig:site}. In Fig.~\ref{fig:overall}, we show the evolution of the current sheet as it develops in the turbulent plasma. The panels show the  out-of-plane current density in a local coordinate system constructed using the method of \citeA{denton:2018}, which will be described in more detail later. Unlike two-dimensional simulations where the current sheet forms along the $z$ (out-of-plane) direction, this current sheet is primarily oriented along the $y$ direction, as shown by the orientation of the planes in (b-d) relative to the simulation axis glyph.  At $t\Omega_{ci} = 17$, there is a region with small positive $j_M$ just below and to the left of the intersection of the axes marked by the arrow. The current density intensifies at $t\Omega_{ci} = 17.25$ as it moves upward, before becoming its most intense in the final panel at $t\Omega_{ci} = 17.5$. We also note there is some out-of-plane motion of the current sheet that is not captured by this figure, but the overall increase of the intensity of $j_M$ in this current sheet holds within the volume around this plane. This process takes place over $0.5/\Omega_{ci}$ using the initial conditions, or approximately $2.5/\Omega_{ci,local}$ using the local magnetic field, meaning the event is transient. 

%At $t\omega_{ci} = 17.5$, the reconnection site is found in the region around $(x,y,z)=(338,188,118)$, where there is a region of enhanced current density, electron flow reversal and a change in magnetic field topology.

In order to characterise the reconnection region, we find a local coordinate system for the current sheet using a hybrid minimum variance/maximum directional derivative method \cite{denton:2018}. In this system, $L$ is the direction of the reconnecting magnetic field, $N$ is normal to the current sheet and $M$ completes the orthogonal triad. For this event, $\vec{\hat{e}}_L = (0.74,0.52,0.41)$, $\vec{\hat{e}}_M = (-0.32,0.82,-0.46)$ and $\vec{\hat{e}}_N = (-0.59,0.22,0.78)$ when evaluated at $t\Omega_{ci} = 17.5$. Due to the motion of the current sheet, the $LMN$ coordinate system does not remain the same with time, though the directions remain similar. For instance, after accounting for the motion of the current sheet, $\vec{\hat{e}}_L$ at $t\Omega_{ci} = 17.25$, which shows the largest change, is rotated approximately $18^\circ$ from $\vec{\hat{e}}_L$ at $t\Omega_{ci} = 17.5$. The coordinate system is illustrated in panels (b) and (c) of Fig.~\ref{fig:site}, in which the $L$, $M$ and $N$ axes are red, yellow and green respectively, and the original Cartesian system is shown by the glyph in the lower left corner. In terms of the simulation coordinates, the current is mainly flowing in the $y$ direction, in contrast to two-dimensional simulations where the reconnection current is restricted to flow in the $z$ direction \cite{bessho:2018shock,bessho:2020}.

The structure of the reconnection region and its signatures can be seen in Fig.~\ref{fig:site}, which shows two-dimensional plots in the $L$-$N$ plane and three-dimensional views from different angles. The colored lines in Fig.~\ref{fig:site}(a-c) are magnetic field lines traced from points close to the current sheet, and are associated with four distinct regions. The black and yellow lines are associated with the reconnection inflows, while the pink and green lines are associated with the outflows. Panel (b) shows the electron outflow velocity in the simulation frame, and it can be seen that there is a flow reversal close to the region where the four different types of field lines meet, indicative of a reconnection region.  Panel (c) shows a volume rendering of the enhancement of current density in the $M$ direction, with values of $J_M$ below 0.28 made transparent. This figure illustrates that the current sheet has a finite extent in the $M$ direction, and we find that its width at the half maximum of $J_M$ is approximately $14$ $d_e$.

\begin{figure}
    \centering
    \ig{3.375in}{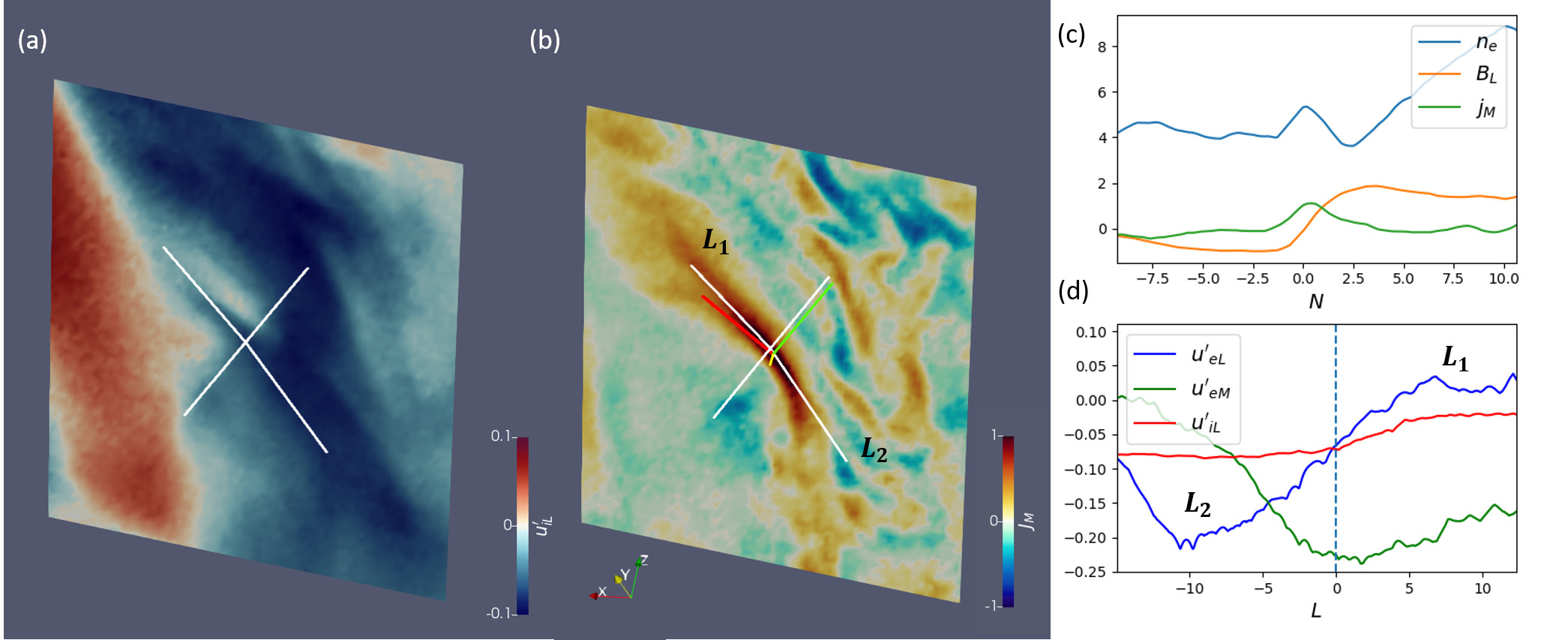}
    \caption{(a) Ion velocity $u_{iL}'$ in the frame moving with the reconnection site. (b) Current density in the $L$-$N$ plane. The white lines show the cuts along which quantities are evaluated in (c) and (d). (c) Density, magnetic field and current density along the $N$ direction. (d) Ion and electron velocities along the $L_1$ (top left)  and $L_2$ (bottom right) cuts. }
    \label{fig:lineplot}
\end{figure}

Further evidence of active reconnection is shown in Fig.~\ref{fig:site}(d-f). Panel (d) shows $\vec{E'}\cdot\vec{J}$ where $\vec{E'} = \vec{E} + \vec{v}_e \times \vec{B}$, indicating the conversion of energy from fields to particles which can occur in reconnection regions. In panels (e) and (f), we use a recently developed diagnostic based on magnetic flux transport (MFT), which has been used to diagnose reconnection sites in turbulent regions in both simulations and observations even when flow patterns are unclear \cite{li:2021,qi:2022}. For this reconnection site, the MFT velocity is calculated as
\begin{equation}
\vec{U}_\psi = \frac{E_M}{B_p}(\vec{\hat{e}}_M \times \vec{\hat{b}}_p),
\end{equation} 
where $E_M$ is the electric field in the $M$ direction and $B_p$ is the in-plane ($L$-$N$ plane) magnetic field. This quantity can be applied to 3D simulations \cite{li:2021} and shows the transport of in-plane magnetic flux. In Fig.~\ref{fig:site}(e,f), the converging inflows of magnetic flux can be seen in the $U_{\psi,N}$ signature, while the $U_{\psi,L}$ signature exhibits diverging outflows. We also note that the guide field for this event is approximately -0.12, approximately 9\% when compared to the upstream magnetic field in the $L$ direction. 

%\begin{figure}[t]
%    \centering
%    \ig{3.375in}{figure_3_mft}
%    \caption{Evidence of active magnetic reconnection. The left and centre panel show the magnetic flux transport velocity $\vec{U}_\psi$, while the right panel shows $\vec{E}' \cdot\vec{J}.$ }
%    \label{fig:mft}
%\end{figure}

To calculate the reconnection rate, we study the asymmetry of the reconnection region. Fig.~\ref{fig:lineplot}(c) shows the density, magnetic field and current density along a cut in the $N$ direction across the current sheet indicated by the white line in panel (b). The current sheet is asymmetric, with $B_1 = 1.8$, $n_1 = 4.3$, $B_2 = 0.95$ and $n_2 = 4.3$, where $n$ and $B$ are measured at the location where $B$ is maximized on either side of the current sheet. The density in the center of the current sheet is higher, and we note that there is density variation in the $L$ direction as well, ranging from approximately $3$ on the left outflow to $7.5$ on the right outflow. The electron and ion $L$ (outflow) velocities are plotted in Fig.~\ref{fig:lineplot}(d), where the primed quantities indicate the velocities are measured in the frame moving with the $B_L$, $B_N$ reversals. Due to the curvature of the current sheet, the outflows are not exactly aligned, as indicated by the lines labeled $L_1$ and $L_2$ in Fig.~\ref{fig:lineplot}(b). As seen in Fig.~\ref{fig:lineplot}(d), the $L_2$ outflow with negative $u_{eL}'$ is stronger. The participation of the ions in this reconnection event is an interesting question. As seen in Figs.~\ref{fig:lineplot}(a) and (c), while bulk ion flow in the $L$ direction is negative in the moving frame, there is a (white) region of less negative ion velocity just above the current sheet. However, this is associated with a region of enhanced ion density outside the current sheet away from the electron flow reversal, and there is no strong ion flow along the $L_2$ cut corresponding to the larger electron outflow.

From asymmetric reconnection scaling laws, the theoretical electron outflow velocity is given by $v_{theory}^2 = B_1B_2(B_1 + B_2)/(\rho_1 B_2 + \rho_2 B_1)$, where $\rho$ is the electron mass density rather than the ion density used for standard asymmetric reconnection \cite{cassak:2007}. Using upstream parameters, the theoretical outflow velocity is 0.64, whereas the peak outflow velocity in this plane is approximately 0.22. If we modify the prediction to account for the pressure variation in the outflow direction \cite{murphy:2010}, the predicted velocity is reduced to 0.53. The reconnection electric field in the region is approximately $E_M = 0.1$, so that the reconnection rate calculated using the theoretical outflow speed $E_M/B_d v_{theory} = 0.15$ where $B_d = 2B_1 B_2/(B_1+B_2)$. When using the measured outflow speed instead, the rate is 0.36. These values are reasonable when compared with reconnection in two-dimensional shock simulations \cite{bessho:2018shock,bessho:2020,bessho:2022}.

\begin{figure}
    \centering
    \ig{3.375in}{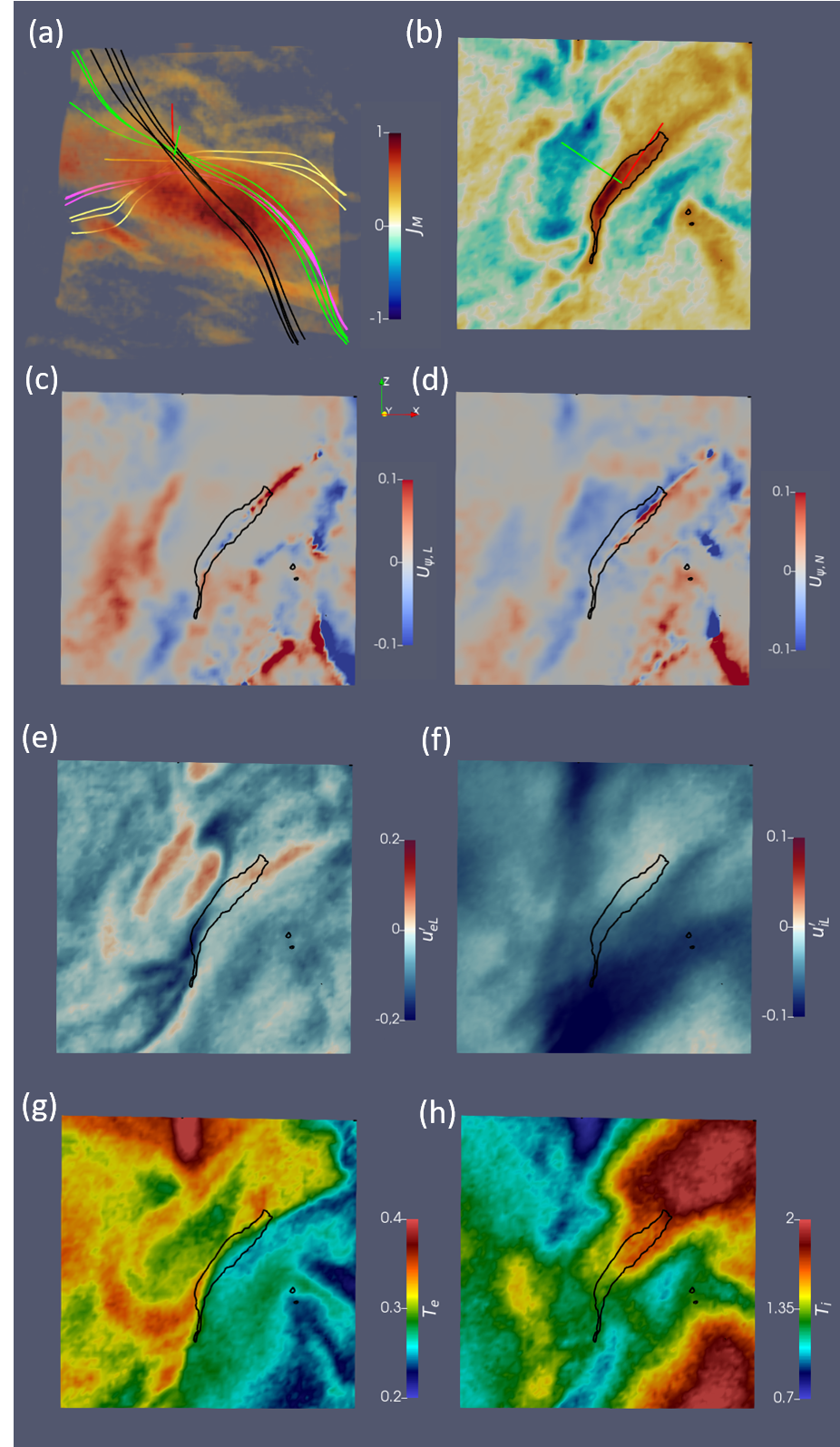}
    \caption{Guide-field reconnection site. (a) Current density $J_M$ and magnetic field lines. (b) Current density $j_M$ in the $L$-$N$ plane. $L$ and $N$ axes are red and green respectively, and the lines have length 10 $d_e$ for scale. (c, d) MFT velocity showing the in- and outflow of magnetic flux, indicating active reconnection. (e) Electron $L$ velocity in the frame moving with the x-line. (f) Ion velocity $L$ in the frame moving with the x-line. (g) Electron temperature. (h) Ion temperature.}
    \label{fig:guidefield}
\end{figure}

To illustrate the different configurations of reconnection, we show a second event in Fig.~\ref{fig:guidefield}. In this case, there is an intermediate $B_M$ of approximately 0.7 compared to the nominal upstream magnetic field $B_d$, indicating that this region exhibits guide-field reconnection. The $L$, $M$, and $N$ axes are illustrated in panels (a) and (b), and the axis directions remain similar over the preceding $0.5/\Omega_{ci}$, similar to the first event. Because of the presence of the guide field, the change in field-line topology is not as drastic as in the previous event, as all the field lines are traced out in the $M$ direction. Instead, one can observe the magnetic shear from one side of the sheet to the other, and the spreading of field lines traced from different regions in the vicinity of the $B_L$-$B_N$ reversal. The current sheet is extended in the $M$ direction with a length of approximately $30$ $d_e$ as shown in panel (a). In the $L$-$N$ plane, the current density $J_M$ is shown in panel (b), where the black line is a contour of $J_M = 0.5 J_{M,max}$ to serve as a guide in subsequent panels.  Signatures of reconnection can be seen in panels (c) and (d), where the $\vec{U}_\psi$ shows inflows and outflows of magnetic flux. The structure of the electron flows is shown in panel (e), and there are oppositely directed flows in the top-right and bottom-left of the region inside the contour. Again, the jets are not collinear or symmetric, with the negative jet reaching a peak of approximately $u'_{eL} = -0.2$ and the positive jet only reaching $u'_{eL} = 0.05$ in the x-line frame. Similar to the previous event, the peak velocity is lower than the theoretical velocity for this event, which is approximately 0.6. As seen in panel (f), there are negative $L$-directed ion flows in the region in and around the current sheet, with the flows most negative in the bottom-left, and least negative towards the top-right, indicatating they may be associated with the reconnection event, though the region is too small for ions to couple fully to the reconnection process \cite{pyakurel:2019}. These ion flows also do not show the same small-scale structure as the electron flows.  
This event shows some electron heating, as seen in panel (g) where there is an increase in electron temperature towards the upper-left of the current layer where density is lower, similar to experimental findings \cite{yoo:2017}. The ion temperature shows variation in both $L$ and $N$ directions at larger spatial scales and may not be related to the reconnection event. The reconnection electric field shows spatial variation and is stronger upstream. Within the current sheet, when  normalized to $B_d v_{out}$, the reconnection rate is approximately 0.05, lower than the first event but within the range of values seen in \citeA{bessho:2022}.

%The maximum measured outflow velocity in this reconnection region is $u_{eL}' = 0.2$, while $E_M = 0.12$. 

% In Ref.~\cite{pyakurel:2021}, it was shown that the reconnection rate in three dimensions can be increased due to additional mass flux in the $M$ direction. If there is net outflow in the $M$ direction, the inflow will have to increase to compensate, leading to an increased reconnection rate compared to a similar two-dimensional geometry. For this event, there is net flux into the region, so the rate would be decreased?

%Notes on MMS event:
%$v_out$ is 4201 km/s, $v_M$ max is 1500, $v_L$ max is 500.

%{\color{red} space for observations or another event.}

\section{Discussion}

The three dimensional reconnection sites show similarities and differences compared to earlier two-dimensional simulations of reconnection in quasi-parallel shocks. Similar to two-dimensional simulations, we find multiple reconnecting and non-reconnecting current sheets, and the structure of reconnection regions is different from laminar reconnection regions such as those in the magnetotail, with asymmetric inflows and outflows \cite{bessho:2018shock,bessho:2020}.

On the other hand, there were differences in the occurrence and regime of reconnection in the 3D simulation. In an equivalent two-dimensional simulation (in the $x$-$y$ plane) we performed with the same physical parameters, we did not find reconnecting current sheets at the same stage of evolution. The additional degree of freedom allows reconnection to occur in different planes. This has more consistency with observations as they show that reconnecting current sheets are observed for a wide range of shock orientations \cite{gingell:2019}.

For the two events discussed in this paper, the orientation of the current sheets are such that $J$ is mainly in the $y$ direction, which would not be possible in a 2D simulation in which the reconnecting current is constrained to be in the $z$ direction. In the $x$-$y$ plane, we see long wavelength waves in the upstream region propagating oblique to the magnetic field with wavelength $\approx 2.5 d_i$, similar to \citeA{bessho:2020}, where oblique waves with wavelength $\approx 3 d_i$ are seen. In the 2D case, these fluctuations cause the bending of field lines which contributes to the occurrence of reconnection. In the $x$-$z$ plane in 3D simulations, the projection of the wavevector of the fluctuations is almost parallel to the in-plane magnetic field, which has an angle of approximately $20^\circ$ to the shock normal. The parallel fluctuations bring alternating $B_z$  to the shock, which may account for the existence of reconnection regions oriented in the $x$-$z$ plane.

 %The range of angles for which simulations of quasi-parallel shocks can be used to study reconnection is then increased, which is important as 

Another major difference between two- and three-dimensional simulations is the ability to study weak guide field reconnection. In two dimensional simulations (e.g.~\cite{bessho:2018shock, lu:2021}), the evolution of the system leads to the generation of strong out-of-plane magnetic fields. Reconnection events observed in such simulations are mainly in the strong guide field regime. In this three-dimensional simulation, we find reconnection events with a range of guide fields, with the first event we focus on having a weak guide field since the orientation is such that the strong $B_z$ contributes to the reconnecting component of the magnetic field. Again, this is relevant to observations, as a statistical study of reconnection in the shock transition region has indicated that reconnection sites have a wide range of guide fields, with stronger guide fields slightly favoured \cite{gingell:2019}.

The two events studied have reconnection rates on the order of 0.1 $B v_{Ae}$, in agreement with other simulation studies of reconnection in shock turbulence \cite{bessho:2022}. No particular enhancement of the reconnection rate for these events due to three-dimensional effects is seen as in \cite{pyakurel:2021}, but a further statistical study will need to be conducted to determine if the reconnection rate in shock-driven reconnection  is modified by 3D effects.

This work was supported by DOE Grant DESC0016278, NSF Grant AGS-1619584, NASA Grants 80NSSC18K1369, 80NSSC20K1312, 80NSSC21K1046. Simulations were performed using NASA HECC resources. 

\bibliography{reconnectionbib}
%\bibliography{reconnectionbib.bib}

\end{document}